\documentclass[aps,prl]{revtex4}
\def\ee{\end{equation}}
\def\bea{\begin{eqnarray}}
\def\eea{\end{eqnarray}}

\def\ul#1{ \underline{#1}}

\def\Prob{{\rm Prob}}
\begin{document}

\title{Path Integrals and Reality}

\author{Adrian \surname{Kent}}
\affiliation{Centre for Quantum Information and Foundations, DAMTP, Centre for
  Mathematical Sciences, University of Cambridge, Wilberforce Road,
  Cambridge, CB3 0WA, U.K.}
\affiliation{Perimeter Institute for Theoretical Physics, 31 Caroline Street North, Waterloo, ON N2L 2Y5, Canada.}
\email{A.P.A.Kent@damtp.cam.ac.uk} 

\date{May 2013} 

\begin{abstract}
We define the idea of {\it real path quantum theory}, a realist
generalisation of quantum theory in which it is postulated that
the configuration space path actually followed by a closed quantum
system is probabilistically chosen.  This is done 
a postulate defining probabilities for paths, which we propose
are determined by an expression involving path amplitudes and
a distance function that quantifies path separation.
We suggest a possible form for a path probability postulate and explore   
possible choices of distance function, including choices
suitable for Lorentz or generally covariant versions of
real path quantum theory.  We set out toy models of quantum
interferometry and show that in these models the probability postulate and
specific distance functions do indeed give a physically sensible
path ontology.  These functions can be chosen so as to predict quantum interference
for interference of microscopic quantum systems and the failure of
interference for macroscopic quantum systems.  More generally, they
predict interference when the beams are close, and its failure when
they are far apart, as determined by the distance function. 
If taken seriously in its present relatively unconstrained form,  
real path quantum theory thus motivates experimental tests of 
quantum interference in all unexplored regimes defined by
potentially physically interesting parameters, including the mass of the
beam object, the beam separation distance, the beam separation time, 
and many others. 
We discuss open questions raised by these ideas. 
\end{abstract}
\maketitle
  
\section{Introduction}

Feynman's path integral formulation  \cite{feynman1965quantum} of quantum theory is widely seen as 
an elegant and beautiful unifying principle that may yet  turn
out to be the fundamentally correct way to define quantum field
theory and quantum gravity.   It also motivates Hartle and Hawking's
intriguing no-boundary proposal \cite{PhysRevD.28.2960} and other
theories of the cosmological initial conditions, giving rise to some
hope of a unified theory of dynamics and boundary conditions. 
Rigorously defining path integrals in physically relevant quantum
theories poses formidable and unresolved technical difficulties.   
Path integrals are nonetheless often said to give an elegant and intuitively 
appealing explanation of the relationship between classical and 
quantum theories, and specifically to allow a simple derivation of the classical principle of
stationary action from quantum theory.   While  
the technical problems in rigorously defining path integrals are generally acknowledged to
be formidable, the conceptual and logical problems in the path
integral account of the relationship between classical and quantum
theories have received, surprisingly,
relatively little attention. 

In this paper we first discuss why, even if we had a
mathematically rigorous path integral for some preferred choice of
variables, we could not use it to explain 
why macroscopic objects approximately follow classical trajectories.

We then explore a new way of understanding quantum
path integrals, defined by a a new path probability postulate
that involves a relatively simple modification to the standard path
integral.   
The proposal has 
some clear conceptual advantages compared to the standard path
integral.   It gives a clear physical meaning to  
the paths and to probabilities associated with them.  
It also suggests a clear and conceptually unproblematic way of 
justifying from first principles the appearance of quasiclassical
trajectories. 

Like the standard quantum path integral, our modified path integrals
are not presently rigorously defined.
However, the formal path integrals we consider do at least tend to suppress contributions
from ``pathological'' paths -- that is, paths that are very far from
intuitively sensible representations of the physics of
the relevant system.  This perhaps offers some grounds for hope that more rigorous definitions
might be achievable, at least for a wider range of physically interesting
models than those that have rigorously defined standard path integrals.  
Another unresolved issue is that our path probability postulate
requires a choice for a distance function
between paths, and at present we see no unique natural choice.
However, some simple and interesting possibilities suggest
themselves.  

We illustrate the ideas of real path quantum theory in toy models.
In these models, the generalized path integral rules we consider
give a simple and physically reasonable ontology. 
In general, they make different experimental 
predictions to those made by standard quantum theory.
These differences can be small enough to be undetectable in microscopic
interference experiments but large enough to predict that macroscopic
objects follow definite classical trajectories, even in experiments where
quantum theory predicts they should display interference.  

We focus throughout on position space path integrals. This follows
a venerable tradition \cite{bohm,debroglie,ghirardi1986unified,GPR}
in being willing to pay the price of singling
out some particular variable or variables in order to define 
equations structures that address the problem of the appearance
of quasiclassicality within quantum theory and the broader quantum reality
problem.  It also follows the mainstream of that tradition in seeing
position as a natural choice.   However, we certainly do not mean
to exclude other choices from consideration; it would be
interesting to explore the various possibilities.   And indeed,
of course, if our ideas can be successfully applied to path integral
formulations of quantum gravity, some more fundamental choice --
perhaps involving paths of geometries -- would need to be made.  

It will be evident that the research program we
outline here is incomplete: this paper begins
to explore ideas and raises some difficult questions
for which we presently do not have answers. 
That said, at the technically unrigorous level of argument used in
standard path integral discussions, our proposed axioms {\it do} suggest
a potentially conceptually satisfactory unified explanation for the quasiclassical
behaviour of macroscopic objects and the observation of quantum
interference in microscopic systems.   They also suggest a way of defining a large class of
generalizations of quantum theory -- {\it real path quantum theories } --
that are equipped with a natural realist ontology and that have
experimentally testable consequences.    
While there are certainly many important and potentially daunting 
unresolved technical issues, 
there seems no intrinsic conceptual or logical obstacle to defining this ontology  
in a way that respects Lorentz invariance, general covariance, and
other symmetries. 

In summary then, we have some new ideas that motivate a research program 
with the ambitious aim of generalising the quantum
path integral to give a unified description of microscopic and
macroscopic physics that is consistent with special and general
relativity and applicable to quantum field theory and quantum gravity.  
We comment briefly on the
relationship of these ideas to other work on finding realist 
versions of standard or generalised quantum theory, at the end of 
the paper.  

\section{Path Integrals and the Principle of Stationary Action} 

\begin{quotation}

``Before we go on making the mathematics more complete, we shall compare
this quantum law with the classical rule. At first sight, from Eq.
(2.15), all paths contribute equally, although their phases vary, so
it is not clear how, in the classical limit, some particular path
becomes most important.  The classical approximation, however,
corresponds to the case that the dimensions, masses, times, etc., are
so large that $S$ is enormous in relation to $\hbar$.  Then the phase
of the contribution $S / \hbar$ is some very, very large angle.  The
real (or imaginary) part of $\phi$ is the cosine (or sine) of this
angle.  This is as likely to be plus as minus.  Now if we move the
path as shown in Fig 2-1 by a small amount $\delta x$, {\it small on
  the classical scale}, the change in $S$ is likewise small on the
classical scale, but not when measured in the tiny units of $\hbar$.
These small changes in path will, generally, make enormous changes in
phase, and our cosine or sine will oscillate exceedingly rapidly
between plus and minus values.  The total contribution will then add
to zero; for if one path makes a positive contribution, another
infinitesimally close (on a classical scale) makes an equal negative
contributin, so that no net contribution arises.

Therefore, no path really needs to be considered if the neighbouring
path has a different action; for the paths in the neighbourhood cancel
out the contribution.  But for the special path $\bar{x} (t)$, for
which $S$ is an extremum, a small change in path produces, in the
first order at least, no change in $S$.  All the contributions from
the paths in this region are nearly in phase, at phase $S_{\rm cl} /
\hbar$, and do not cancel out.  Therefore, only for paths in the
vicinity of $\bar{x} (t)$ can we get important contributions, and in
the classical limit we need only consider this particular trajectory
as being of importance.  {\bf In this way the classical laws of motion
arise from the quantum laws.}

We may note that trajectories which differ from $\bar{x} (t) $
contribute as long as the action is still within about $\hbar$ of
$S_{\rm cl}$.  The classical trajectory is indefinite to this slight
extent, and this rule serves as a measure of the limitations of the
precision of the classically defined trajectory.''

(Feynman and Hibbs, \cite{feynman1965quantum}; italics original, bold face added) 

\end{quotation}

Some version of this argument is still propagated in many quantum 
theory lecture courses to this day.  However, it is conceptually 
and logically confused, and the main conclusion -- which we have highlighted
in bold face above -- does not follow.  

\subsection{Decoherence and the appearance of quasiclassicality: the Feynman-Hibbs lacuna in context} 

Feynman and Hibbs start from a true
statement -- or, at least,
one that could be true if rigorous definitions were available --
that notes that two different situations are related {\it
mathematically}. 
Namely, we can calculate
{\it as if} the approximately classical trajectories were the
only relevant ones, even though actually they are not.  
They add the premise that quantum theory 
is fundamentally correct -- which superficially may seem reasonable enough, 
since we don't have a better theory.
They also add the empirical observation that we
see classical systems approximately following classical equations
of motion.   But then, rather than testing whether we can actually {\it derive} a  
fully consistent explanation of the appearance of quasi-classical
physics from quantum theory, they effectively {\it assume} the answer. 
That is, they assume the appearance of quasiclassicality {\it must} be directly derivable
from path integral quantum theory.  Given that, the calculational result {\it must} be a
derivation of the empirical observation, since it is essentially the only relevant equation that
the theory gives us.   But, of course. 
being able to calculate {\it as if} something were true isn't the
same as showing that it is true from first principles.  

Another version of this error arises in attempts to explain
the appearance of a quasiclassical world from quantum decoherence,
starting from the correct observation that decoherence models
show that one can calculate {\it as if} an initially pure quantum system 
is represented by a proper probabilistic mixed state after
interacting with an apparatus, although in fact it is represented
by the reduced density matrix of an entangled state.    
Extended discussions of this point can be found in
Refs. \cite{giulini1996decoherence,mwbook} and elsewhere.  

Not only do these attempts to derive the appearance of 
quasiclassical world from unitary quantum theory fail, but
-- in the view of many physicists -- all such attempts have failed.
Unitary quantum theory can only be
made sense of via many-worlds ideas, and there is, after 
$55$ years, no consensus even among proponents of those
ideas as to how they can be made rigorous and can give
a scientific theory with explanatory power \cite{mwbook}. 

This motivates exploring ways to go beyond standard quantum theory,
for example by adding extra mathematical structure (as in 
de Broglie-Bohm theory \cite{bohm,debroglie}) or new dynamical laws (as in GRWP
models \cite{ghirardi1986unified,GPR}).  We explore a new idea in this direction below, 
adapting the existing path integral formalism by adding new postulates. 

That said, as we have indicated, 
there is of course not currently a complete consensus on whether quantum theory can
explain the appearance of quasiclassicality.  
For those still unpersuaded that it cannot, a more conservative motivation is that, whatever the status 
of standard quantum theory, natural-looking additions or alterations
should be explored, since they might either give
a valuable new perspective on standard quantum theory, or interesting
new generalizations of quantum theory that can be tested.    

\subsection{Examining the Feynman-Hibbs argument in a toy model}


To see the problem with the Feynman-Hibbs argument more clearly, it is very helpful to separate 
conceptual questions from the problems of rigorously defining 
any path integral. 
To this end we define a toy discrete path integral
model (which we will call $M1$) for the centre of mass motion of a 
single massive object in position space, 
involving some large finite number of paths from 
$A$ to $B$.   Because the number of paths is finite, we can rigorously
define the path integral and related quantities.  We define the model
to have a set of paths that have mathematical properties analogous to
those of quasiclassical trajectories -- those in the neighbourhood of the 
stationary action path --  in the standard quantum path
integral.   This allows us to focus
on the conceptual question of what conclusions about quasiclassical
physics do or do not follow from the path integral.     

For simplicity, we define the 
model $M1$ so that each path has phase $\pm 1$, and we take the paths to have
some natural ordering $P_1 , \ldots , P_N$, in which paths $P_i$
and $P_{i+1}$ are supposed to be physically adjacent.   
This is not generally entirely realistic, even in the simple versions
of path integrals defined in discrete models of $1+1$ dimensional
space-time.
However, it simplifies the model while still allowing it to illustrate a key point that can
be replicated in more geometrically realistic models. 

We suppose also that we can 
identify paths $P_M , \ldots , P_{M+K}$ that correspond
approximately to the quasiclassical trajectories for the particle, where $M$ is 
odd, $N-M-K$ is even, $1 < M < M+K < N$ and $K \ll M,N$.  
We call these the {\it quasiclassical paths} in the model. 
Finally, we suppose that the path amplitudes $A(P_i)$ obey
\begin{eqnarray} 
A(P_i ) &=& (-1)^{i-1} {\rm~for~} 1 \leq i \leq M-1 \, , \\
A(P_i) &=& 1 {\rm~for~} M \leq i \leq M+K \, , \nonumber \\
A(P_i) &=& (-1)^{i-M-K} {\rm~for~} M+K < i \leq N \, . \nonumber 
\end{eqnarray} 
In other words, the amplitudes alternate in pairs before and after
the quasiclassical paths, which all have amplitude $+1$.
Listed in order they are  
$$1, -1 , \ldots , 1 , -1 , 1 , 1, \ldots 1, -1 , 1 ,
\ldots , -1 , 1 \, . $$

This is intended to model the features of the quantum path integral
relevant to Feynman and Hibbs' argument.  The quasiclassical paths
in the quantum path integral are close to one another; in our discrete
model they are all adjacent.   The quasiclassical path amplitudes 
in the quantum path integral are approximately constant; in our model
they are precisely equal, taking the value $+1$.   The amplitudes 
of paths away from the quasiclassical paths oscillate rapidly in the quantum  
path integral; in our discrete model they oscillate as rapidly as
possible, alternately taking the values $\pm 1$.   


Since the path amplitudes either side of the quasiclassical
paths cancel in pairs, we have the arithmetical identity 
$$
\sum_{i=1}^N A(P_i ) = \sum_{i=M}^{M+K} A( P_i ) \,  .
$$ 
So we can indeed calculate the total sum -- our discrete version of the 
path integral -- by summing the amplitudes of the quasiclassical
paths and ignoring the rest.  
But notice that this property {\it per se}
does not single out the quasiclassical paths in our model as special.    
For example, we could also write 
$$
\sum_{i=1}^N A(P_i ) =  \sum_{i=1}^{K+1} A(P_{2i-1}) \, . 
$$
More generally, 
$$
\sum_{i=1}^N A(P_i ) = \sum_{i \in I} A( P_i ) \, 
$$
for any size $(K+1)$ subset $I$ of the set 
$$\{1, 3, \ldots , M, M+1 , \ldots , M+K , M+K+2
, \ldots , N \} = \{ i : A(P_i) = + 1 \} \, , $$
that is, the set of paths with amplitude $+1$.  

It is true that the quasiclassical paths are all adjacent in our ordering, while the other paths are
not.   But nothing in the definition of the path integral or any
standard presentation of its physical implications gives a 
special {\it ontological} status to subsets of adjacent paths.
To derive classical laws of motion, we need to be able
to make a statement about the actual trajectories of macroscopic
objects.  In particular, here, we need to be able to derive that the
object follows one of the quasiclassical 
paths from $A$ to $B$.  This does not follow from the rules of standard path integral quantum theory,
as set out in Feynman and Hibbs or elsewhere. 

This point is worth elaborating.  The standard treatment of the quantum path
integral only defines a transition probability from $A$ to $B$. 
It does not supply 
a rule that tells us that the system actually follows {\it any} path. 
In particular, it gives no rule that ensures the system will
follow one path from among a set of adjacent paths with
similar phases and amplitudes, or even that we can make some more 
coarse-grained statement about its behaviour characterized by that
set.  

In fact, when path integrals are being 
discussed in contexts where no quasiclassical dynamics is expected
to emerge, authors often suggest an intuitive picture according to
which, since every path amplitude has modulus one, in some loose sense they are all 
equally significant: the quantum system ``follows all possible
paths''.   

Of course, this intuition isn't properly justified {\it
either}.   It's not even clear what it is really intended to mean. 
Sometimes some form of Everettian many-worlds picture seems to be
intended.  While it is difficult to criticize so underdeveloped an
intuition, three comments are worth making here. 
First, there have been many different attempts to define an ontologically
sensible and scientifically useful Everettian picture of quantum
theory \cite{mwbook}, and in the view of many 
(e.g. \cite{kent1990against,kentoneworld,albert,price}), none of them succeed. 
Second, elementary paths in the path integral cannot be identified
with the quasiclassical branching worlds that are normally thought
to be crucial in Everettian explanations.     
Third, in any case, without an ontological rule, we can't use the path integral to say
{\it anything} about what the system does between $A$ and $B$. 

Notice also that this intuition about the meaning of the path
integral for microscopic quantum systems directly conflicts with the intuition
discussed above for macroscopic objects following quasiclassical trajectories. 
According to this folk intuition, they {\it don't} follow all possible
paths, but instead follow some quasiclassical path.   

As this conflict of intuitions highlights, it is not the case that
there is some tacit rule, generally familiar to experts but
unaccountably omitted from textbooks, that unifies the path integral
treatments of the microscopic and the macroscopic.   
We have here a genuine conceptual problem that needs to be resolved.  

Fortified by the inescapability of this conclusion, while also 
admiring the beauty and generality of the path integral formalism and suspecting that
despite its present flaws the Feynman-Hibbs argument may contain the germ of a key insight
about the relationship between quasiclassical and quantum physics, 
we now look for fresh inspiration in the form
of alternative ways of thinking about quantum path integrals that might make more  
conceptual and physical sense.    

\section{Real Path Quantum Theory}
  
\subsection{Unphysicality of path probabilities in the standard path
  integral}

Consider the position space path integral for a single
non-relativistic particle transition from point $A= ( x_A , t_A )$
to point $B = (x_B , t_B )$.  
Assigning probabilities to individual paths in this integral makes
no evident sense, as noted above. 
Consider again the naive
rule that the probability of following any given path $P$ is
proportional to the square of the associated amplitude: $\Prob (P) = C | A(P) |^2 $.
Each path $P$
has amplitude $A(P) = \exp ( i S(P ))$, so adopting this rule would make all paths
have equal probability weight. 
In realistic models, these weights are unnormalisable, so that we
cannot obtain a path probability distribution from this rule.
Nor, even if we could somehow solve this problem,
would adopting this rule help explain the origins of quasiclassicality, 
since if any sensible definition of measure existed,
the approximately classical paths of a macroscopic object should 
have measure zero among the set of all paths.  


\subsection{The path probability postulate}

If we can't extract a sensible explanation of quasiclassical physics 
(or indeed do anything more than calculate transition probabilities)
from the path integral in its present form, 
then perhaps we need to change the definition of the path integral, or
add additional postulates, or both.   
The difficulty in making physical sense of the path integral 
seems to be connected with the fact that it hints at an
interpretation in which paths have probabilities, while
at the same time suggesting conflicting intuitions about these
physical path probabilities.
We thus propose to explore the implications of explicitly
assigning probabilities to paths via a new postulate. 
The idea here is that we define 
a quantity $\Prob (P)$ that has the standard properties of 
a probability:
$$
\Prob (P) \geq 0 \qquad \, \int_{{\rm paths~}P} dP \,  \Prob (P) = 1 \, . 
$$
This quantity represents the probability that the given path $P$ 
was actually followed.   Physical reality -- in an experiment, or,
in principle, in the evolution of the cosmos from initial to final
state -- is given by the chosen path.   
 
Specifically, we will consider a postulate of the form:
\begin{equation}\label{pathprob}
\Prob (P) = C \, | \int dQ \exp (- i S(Q) )  \exp ( - d (P, Q))
|^2  \, ( \int dQ \exp ( - d (P, Q) ))^{-1}  \, . 
\end{equation}
Here and below we take $\hbar = 1$. 
For the moment we take the integrals in this expression
to be over all paths $Q$ that have the same endpoints
($A$ and $B$) as $P$.  (Note that we would need to allow larger classes of paths
to obtain an effective description of experiments with an 
extended initial wave function or to discuss the general
possibilities allowed in cosmology.) 
We take $d(P,Q)$ to be some distance measure defined between paths $P$ and
$Q$.  This measure $d$ is supposed in
some natural sense (to be elaborated) to say how distinct the paths
are.   

Note that at present we have no compelling reason to believe that there is a 
unique physically sensible choice for either the form
of (\ref{pathprob}) or the distance function $d$. 
We aim to show that the simple path probability rule (\ref{pathprob}) does lead to physically
interesting conclusions for some choices of distance function.
We find this encouraging, since it is not a priori obvious that
there are any modifications of the quantum path integral that
give physically sensible results in agreement with empirical
evidence for both quasiclassical and microscopic quantum systems.
For concreteness, we focus on (\ref{pathprob}) here, and explore
various distance functions.   However, it is certainly also
interesting to explore the range of physically sensible alternatives to
(\ref{pathprob}). 

In the form just given, the path probability postulate assigns probabilities for paths between
$A$ and $B$ conditioned on the fact that $A$ and $B$ are the initial
and final states respectively.   Under this conditioning assumption,
the normalisation factor $C$ is defined by 
\begin{equation}
\int_{{\rm paths:} A \rightarrow B} dP \, \Prob (P) =1 \, . 
\end{equation} 
Without the conditioning assumption on the final state, the
path probability postulate also implies a new rule for the 
total probability for arriving at the final point $B$ from the
initial point $A$:
\begin{equation}
\Prob (B | A ) = \int_{{\rm paths:} A \rightarrow B} dP \, \Prob (P) \, . 
\end{equation}
In this context $C$ is defined by normalising over a complete
basis of final states: 
\begin{equation}
\int_{{\rm final~states~} B} dB \, \Prob (B | A ) = 1 \, . 
\end{equation}
For example, for a single particle in Minkowski space, the 
basis $B$ could be taken to be the final position $\ul{x}$
on any given spacelike hypersurface intersecting the future
light cone of the initial position.   

We now consider what constraints our 
notion of naturality could imply on $d$. 
If $d$ were to define a metric on the space of paths it would satisfy 
the following conditions:
\begin{enumerate}
\item{} $d(P,Q) \geq 0$ for all $P$ and
$Q$ (non-negativity), 
\item{} $ d(Q,P) = d(Q,P)$ (symmetry), 
\item{} $d(P,Q) = 0$ if and only if $P=Q$ (identity of indiscernibles),
\item{} $d(P,R) \leq d(P,Q) + d(Q,R)$ (triangle inequality).
\end{enumerate}
Of these, only non-negativity is strictly needed for our
present discussion.  The symmetry postulate also seems 
very natural at first sight, but there turns out to be 
some motivation to consider asymmetric distances when
considering paths in Minkowski space or other fixed
background Lorentzian spacetimes.  
We will not adopt the identity of
indiscernibles as an axiom here for two reasons. 
First, it is useful in simple toy models to allow
distinct neighbouring paths to have zero separation rather than
very small separation. 
Second, some interesting candidates for Lorentz invariant distance
functions between paths in Minkowski space have the property
$d(P,P)>0$ for some non-causal paths $P$. 
The distance functions in our toy models 
also violate the triangle inequality.  While this too seems 
a plausible candidate postulate for a fundamental theory, 
it is also easy to find simple distance measures that violate it.
We thus do not impose the last three postulates as axioms at
present, but keep them in mind as natural possibilities to 
consider adopting in a fundamental formulation of a  probabilistic path theory. 

In summary, then, in what follows below, $d$ is supposed to give some
intuitively sensible measure of
path separation, and to be non-negative, but is not necessarily a metric. 

We also require a physically motivated condition on $d$: 
that $d(P,Q) \approx 0$ when the difference between $P$ and $Q$
is ``microscopic'' and $d(P,Q) \gg 1$ when the difference
is ``macroscopic''.   The motivating idea here is that the path distance
$d$ is ultimately defined by a new fundamental theory generalising 
quantum theory, and that this definition of $d$ 
is what ultimately allows the theory to determine the boundary between the
``microscopic'' and ``macroscopic''.  

If we are to produce a generalization of quantum theory that has not
already been falsified, 
$d$ should be chosen so that the predictions
are consistent with experiment and observation to date.  
So, paths $P$ and $Q$ in experiments that demonstrate the path
interference predicted by standard quantum theory should be microscopically
separated: $d(P,Q) \approx 0$.   
However, if we see an object following an approximately classical 
trajectory, consistent with a path $P$, and there is another path
$Q$ describing a trajectory that we can distinguish from $P$ by
observation, so that we also see that the object does not follow $Q$,
then $P$ and $Q$ should be macroscopically separated: $d(P,Q) \gg 1$. 

In a truly fundamental formulation of a new theory, the probability rule should apply
to paths between possible initial and final states of the universe,
and so should presumably be formulated within some quantum theory of
gravity.   We will be rather less ambitious initially in exploring the idea,
by making various simplifying assumptions.   

First, we will suppose the probability rule makes sense for paths in the
appropriate configuration spaces in quantum mechanics or quantum
field theory, for finite time intervals or between finitely separated
space-like hypersurfaces.   This does not necessarily 
conflict with the idea that the fundamental formulation
should be for paths between initial and final cosmological states.
The intuition, rather, is that the restricted 
application of the probability rule should be derivable as an
approximation from the fundamental version.   
Similarly, the intuition is that the application to quantum mechanics
should be derivable from that to quantum field theory, which in turn
should be derivable from that to some underlying unified theory 
that includes gravity. 

Second, we simplify considerably further by considering 
discrete toy model versions of path integrals, in which there
are only finitely many relevant paths.   
In the simplest toy models we label these paths numerically, with a
distance function depending on the abstract label, rather than
defining an underlying path geometry and a distance function based
on that geometry.    
We assign the phases of the paths in these toy models
by fiat rather than deriving them from a specified Lagrangian and
action.  
As in our earlier toy model, these phases
are chosen so as to mimic the
essential features of the sort of phase distribution 
one might expect from a realistic path integral calculation, but
are not directly derived from any path integral. 
Our aim is to illustrate the possibility of extracting new physical
insight from the path probability postulate (\ref{pathprob}) for some
reasonably sensible choices of the distance function
$d(P,Q)$, without worrying about questions of rigour.  
We leave for future exploration the scope for
physically sensible choices of $d(P,Q)$ for which
the path probability postulate may be rigorously defined in realistic models.  

\section{Real path quantum theory in toy models}

In principle, the path probability postulate is intended to be an interesting
possibility to explore in any physical theory, including quantum
gravity theories with no fixed background space-time. 
In the first instance, though, we are mainly 
interested in exploring the cases with a fixed background 
space-time.  In the non-relativistic case, we focus here on
the example of Galilean spacetime with trivial topology $R \times R^3$.
In the relativistic case, we focus here on 
Minkowski space and on other fixed background Lorentzian space-times
with the same topology.   Of course, more general space-times with
non-trivial topologies are
also interesting.   We focus on these examples because they are 
sufficient to illustrate the generality of real path quantum theory
ideas, without complicating the discussion by considering non-trivial
topologies.     

The toy models we consier next are abstract enough that they could
apply to each of these cases and to others.   To be concrete, it may be
helpful to think of them as models in Galilean space-time, and
so we will assume this for now.   
We make some comments later about possible choices of distance
function that give potentially physically interesting versions of
the path probability postulate in Minkowski and other Lorentzian space-times.  

\subsection{Modelling single particle beams: Real path quantum theory in the toy model M1} 
 
We now return to our toy path integral model $M1$ above, and develop it as a model
for real path quantum theory, by adding a choice of distance
function. 
Recall that we previously proposed $M1$ as a discrete model of paths of the centre
of mass motion of a massive object between
two specified points $A$ and $B$ in space-time, where $B$ is in the
causal future of $A$.  
Intuitively, we expect such an object approximately to follow
the least action path from $A$ to $B$, even though we cannot rigorously
justify this intuition from within standard quantum theory. 

The same model $M1$ can also be thought of as a model for 
a single quantum particle in a beam between 
a source $A$ and a detector $B$ that -- 
in the ordinary intuitive
but unrigorous language generally used about particle beams in
experiments --  approximately defines a single definite
path from $A$ to $B$, 

We will use the model to consider both cases.
In either case, the point is to show that, if we apply real path quantum
theory to the model, we {\it can} rigorously justify the intuitive 
expectations.    


For the moment we are interested in calculating path probabilities 
conditioned on fixed initial and final points, $A$ and $B$. 
As above, then, we will suppose there
are $N$ possible paths between $A$ and $B$, where $N$ is a large
positive integer, the paths are 
labelled $P_j$ for $1 \leq j \leq N$, and they have corresponding
phases $S_j = S(P_j)$ and amplitudes $A_j = \exp ( - i S_j )$.  The label $j$
is supposed to correspond to the geometric location of the path in space-time,
in such a way that paths $P_i , P_{i+1}$ with adjacent labels are in some sense
neighbouring, and the difference between labels is a measure of the 
separation between paths, and so we will assume that
$d(P_i , P_j ) = f ( | i - j | ) $ for some function $f$.   
We note again that this is not entirely realistic.   If we 
think of the paths as living in some discretized Galilean or Minkowski
space, these properties do not generally hold for interesting 
sets of paths if we use choices of the distance function
that are naturally defined via the underlying geometry.   
However, it simplifies our model and 
allows us to derive some interesting features.
The essential points we make also hold in more 
geometrically realistic models.  

We again suppose some set of adjacent paths $P_M , \ldots , P_{M+K}$ lie in 
a region in path space where the path phase is essentially
constant, while for the remaining paths the path phase
oscillates. 
Here $M$ is 
odd and we take $N-M-K$ to be even (although this is inessential), $1 < M < M+K < N$ and $K \ll N$.  
We again take the 
path amplitudes in the constant region to all be $+1$, and
take the amplitudes for the paths outside the 
region to be alternately $\pm 1$.
So we have 
\begin{eqnarray}
A(P_i ) &=& (-1)^{i-1}  \qquad {\rm~for~}  1 \leq i \leq M-1 \, , \\
A(P_i) &=& 1 \qquad {\rm~for~} M \leq i \leq M+K \, , \nonumber \\
A(P_i) &=& (-1)^{i-M-K} \qquad {\rm~for~} M+K < i \leq N \, \, , \nonumber 
\end{eqnarray}
and thus the path amplitudes listed in order from $A_1$ to $A_N$ are  
$$1, -1 , \ldots , 1 , -1 , 1 , 1, \ldots 1, -1 , 1 ,
\ldots , -1 , 1 \, . $$

We now take the distance function to be 
\begin{eqnarray}
d(P_i , P_j) &=& 0 {\rm~if~} | i - j | < D \, , \\
\qquad d(P_i , P_j ) &=& \infty {\rm~if~} |i-j| > D \, ,  \nonumber \\
\qquad d(P_i , P_j ) &=& \log (1/2) {\rm ~ if ~} | i - j | = D \, . \nonumber 
\end{eqnarray}
This infinite step function
is meant as a simplifying approximation to something more natural,
such as 
$$d(P_i , P_j ) = \exp ( |i-j | /  D ) \, . $$
Here we require $  D \ll N$, $M > 2D+1$, and $N-M-K > 2D+1$.  

Since we are now working within real path quantum theory, we can apply
(\ref{pathprob}) to obtain an explicit expression for the probability of
the particle following any given path:
\begin{equation}
\Prob (P_i ) = C ( 2 D  )^{-1} | \, ( \sum_{|j-i| < D } A(P_j) + \sum_{|j-i|
  = D} \frac{1}{2} A(P_j) ) \, |^2 \, , 
\end{equation}
where $C$ is the normalisation factor ensuring that 
$$
\sum_{i=1}^N \Prob (P_i
) = 1 \, . 
$$

\subsection{Modelling a single particle beam} 

If we think of $M1$ as modelling a single microscopic particle beam between
source $A$ and detector $B$, the most immediately interesting case for us is $2D>K$ (in
particular $2D \gg K$, though the calculations depend only on
whether $2D$ or $K$ is larger).  This parameter choice
captures the intuition that any model that alters the predictions
of quantum theory by introducing some intrinsic decoherence
should ensure that paths lying within a single beam of a
microscopic quantum particle are very far from decohering.
(We see no decoherence for microscopic particles 
even in interference experiments involving
multiple separated beams -- a scenario we will model later.) 

For $2D>K$, and for $i$ further than $D$ from the ends of the list of paths, 
i.e. in the range $D <i< N-D$, 
this gives 
\begin{eqnarray} 
\Prob (P_i ) &=& 0   \qquad {\rm~if~} D+1 < i< M-D {\rm~or~} N- (D+1)> i> M+ K + D \, , \\
\Prob (P_i ) &=& C (2D)^{-1}  | K |^2  \qquad {\rm~if~} M+K-D < i < M+ D \, , \nonumber \\
\Prob (P_i ) &=& C (2D)^{-1} | (i+D-M ) |^2 \qquad {\rm~if~} M < i+D< M+K \, , \nonumber \\
\Prob (P_i) &=&  C( 2D )^{-1} | (M+K - i +D ) |^2  \qquad {\rm~if~} M< i-D<M+K \, . \nonumber 
\end{eqnarray} 

For completeness we also consider the case $2D \leq K$. 
Here, again for paths $P_i$ away from the ends of list, with $i$ lying in the range $D<i<N-D$, we find
\begin{eqnarray}
\Prob (P_i ) &=& 0   \qquad {\rm~if~} D+1 \leq i< M-D {\rm~or~} N -
(D+1) \geq i> M+ K + D \, , \\
\Prob (P_i ) &=& C (2D)^{-1} | 2D |^2 = 2 C D \qquad  {\rm~if~} M+D < i < M+K - D \, , \nonumber \\
\Prob (P_i ) &=& C (2D)^{-1} | (i+D-M ) |^2 \qquad {\rm~if~} i-D<M<i+D \, , \nonumber \\
\Prob (P_i) &=&  C( 2D )^{-1} | (M+K - i +D ) |^2  \qquad {\rm~if~} i-D<M+K<i+D \, . \nonumber
\end{eqnarray}

So, in either case, aside from the paths near the ends of the path
list, all paths with non-zero probability 
are close to the paths in the region of constant phase, in the sense that their index $i$
is within $D$ of that region. 
Paths closer to the region are likelier, while the
probabilities fall off towards zero for paths further away.  

In both cases boundary effects mean our model also gives slightly nonzero path probabilities
for paths near the ends of the path list: 
\begin{eqnarray}
\Prob (P_i) &=& C 
(1/4) ( i + D - \frac{1}{2} )^{-1} \qquad {\rm~ for~} 1 \leq i \leq D \, , \\
\Prob (P_i) &=& C 
(1/4) ( (N-i) + D - \frac{1}{2} )^{-1} \qquad  {\rm~ for~} N-D \leq i \leq N \,
. \nonumber 
\end{eqnarray}
These are artefacts, which would be eliminated if we 
took the path list to be infinite or imposed periodic boundary
conditions; we ignore them as irrelevant to our discussion.    

This toy model thus gives a realist ontology that tells us that, if
the particle goes from 
$A$ to $B$, then it follows a definite path.
It respects physical intuition, in 
that the realised path
will be close to the region of constant phase in path space.  
How close depends on the parameter 
$D$ that characterizes our chosen distance function in this model.

Although our realist ontology is not part of standard quantum theory, 
its dependence on $K$ in this model is consistent with
standard intuitions.  Our parameter $K$ here models (in a loose intuitive
sense, since we are not considering measures on the set of realistic 
paths here) the size of the
set of paths around the stationary path for which   
$(S/ \hbar)$ is approximately constant in standard path integral quantum theory 
quantum theory.    The Feynman-Hibbs argument
discussed earlier also aims to select a set of roughly equally
relevant paths of similar phase around the stationary point of the
action. 

However, a new feature of our models is that the ontology and hence the physical predictions also depend on
the distance function $d$ -- in this case via the parameter $D$.   
The physically relevant paths in our ontology -- those with nonzero
probability -- are not only those in the stationary phase region, but
also those $d$-close to that region.
A more fundamental new feature of our models, of course, is that they we {\it have} a realist
ontology: as already noted, the standard Feynman-Hibbs intuitions have
no logical justification in standard path integral quantum theory.  
 
\subsection{Modelling classical and quasiclassical trajectories} 

As we noted above, the same toy model can be used to describe the centre of mass motion
of a macroscopic object, which we expect to follow a quasiclassical
trajectory.   Whether we should take $2D>K$ or $2D  < K$ here is 
less clear, since we do not have strong experimental constraints
on quantum interference of general macroscopic objects.  
Since both ranges give qualitatively similar results when modelling
a single beam, this does not immediately matter.  
As shown above, our model suggests that, ignoring boundary artefacts,
all the trajectories with non-zero probability 
are approximately quasiclassical, in the (newly defined) sense that their index $i$
is within $D+ \frac{K}{2}$ of the path at the centre of the
quasiclassical set (which we take to represent the stationary action
path
in our model). 
Trajectories closer to the quasiclassical set are likelier, and the
probabilities tail off towards zero for paths further away.  

Our realist ontology thus says that, if 
a classical object's centre of mass goes from 
$A$ to $B$, then it follows a definite trajectory, and this 
trajectory will be approximately the classical one.  How good
the approximation is depends on the size $K$ of the set of 
paths close enough to the classical trajectory that their 
phase is essentially the same as its, and on the parameter 
$D$ that characterizes our distance function.
As already noted, the $K$-dependence is in line with intuitions based on
standard quantum theory.   In particular, it supports the conclusions
of the Feynman-Hibbs argument in the context where it originally was
meant to apply, namely selecting a set of roughly equally
relevant paths of similar phase around a classical trajectory
that is (necessarily) a stationary point of the action. 
Again, we have the new feature of dependence on
the distance function $d$, parameterised in our model by $D$. 

\subsection{Modelling beam interferometry in real path quantum theory} 

We next consider a toy model, which we call $M2$, in which there are two different regions in
path space where the path phase is constant.   Since we want to 
model quantum interference of beams with general complex phases we now let the path amplitudes in these
regions take any complex value of modulus one.  However, to keep the
model simple we still for the moment suppose the amplitudes outside the regions are 
alternately $\pm 1$.
We now have  

\begin{eqnarray}
A(P_i ) &=& (-1)^{i-1} {\rm~for~} 1 \leq i \leq M_0 -1 \, , \\
A(P_i) &=& \exp ( - i \theta_0 ) {\rm~for~} M_0 \leq i \leq M_0 +K_0 \, , \nonumber \\
A(P_i) &=& (-1)^{i-M_0 -K_0 } {\rm~for~} M_0 +K_0 < i \leq M_1 -1 \, \, , \nonumber \\
A(P_i) &=& \exp ( - i \theta_1 ) {\rm~for~} M_1 \leq i \leq M_1 +K_1 \, , \nonumber \\
A(P_i) &=& (-1)^{i-M_1 -K_1} {\rm~for~} M_1 +K_1 < i \leq N \, \, , \nonumber 
\end{eqnarray} 
where we take $M_0$ to be odd, $M_1 -M_0-K_0$ to be odd, 
$N-M_1 -K_1$ to be even, and assume the inequalities
$1 < M_0 < M_0+K_0 < M_1 < M_1 + K_1 < N$, $K_0  \ll N$,
$ K_1  \ll N$,
$M_0 > 2D+1$ and $N-M_1-K_1 > 2D+1$. 
The path amplitudes listed in order from $A_1$ to $A_N$ are
thus now 
\begin{eqnarray}
1, -1 , & \ldots & , 1 , -1 , \exp ( - i \theta_0 )  , \exp ( - i \theta_0 ) , \ldots \exp ( - i \theta_0 ) , 1 , -1 ,
\ldots , \\
& \ldots & , 1 , -1 ,  \exp ( - i \theta_1 )  , \exp ( - i \theta_1 ) ,
\ldots \exp ( - i \theta_1 ) , 1 , -1 ,
\ldots , 1, -1 \, . \nonumber 
\end{eqnarray} 

The most interesting cases for our discussion are (i) $D \gg K_0 , K_1$ and
$D \gg M_1 + K_1 - M_0 $ and (ii) $D \gg K_0 , K_1$ and $2D+1< M_1 -
M_0 - K_0$.   

In case (i), if $i+D > M_1 + K_1$ and $i-D<M_0$ we have
\begin{equation}
\Prob (P_i ) \approx C | ( (K_0 + 1) \exp ( - i \theta_0 )  + (K_1 + 1)
\exp ( - i \theta_1 ) ) |^2  ( 2 D )^{-1}   \, .
\end{equation}
In other words, these path probabilities are defined by an
interference term, which is essentially the term that would arise from two beams
with respective amplitudes $( K_0 +1 ) \exp ( - i \theta_0 )$ and 
$( K_1 + 1) \exp ( - i \theta_1 )$.  

For $i$ in the range where $i-D < M_0$ and $M_0 \leq i+D < M_1 + K_1$,
the paths $P_i$ also have significantly nonzero probabilities
representing partial interference; similarly for $i$ in the range
where
$i+D > M_1 +K_1$ and $M_0 \leq i-D \leq M_1 + K_1$. 

As in the previous model, we also have slightly nonzero probabilities for paths near
the ends of the list, which we again ignore as artefacts.  
   
In case (ii), if $ i -D < M_0 $ and $i+D > M_0 + K_0$, we have
\begin{equation}
\Prob (P_i ) \approx C | ( (K_0 + 1) \exp ( - i \theta_0 )  ) |^2  ( 2
D )^{-1} = C ( K_0 + 1 )^2  (2 D)^{-1}  \, .
\end{equation}
Similarly, if $ i -D < M_1 $ and $i+D > M_1 + K_1$, we have
\begin{equation}
\Prob (P_i ) \approx C | ( (K_1 + 1) \exp ( - i \theta_1 )  ) |^2  ( 2
D )^{-1} = C ( K_1 + 1 )^2  (2 D)^{-1}  \, .
\end{equation}
Outside these ranges, ignoring boundary artefacts, the path
probabilities
fall to zero.  

So, in case (ii), paths with significantly nonzero probability
are associated with (and $d$-close to) either the region of paths
with constant amplitude $\exp ( - i \theta_0 )$ or the region of paths
with constant amplitude $\exp ( - i \theta_1 )$.   However, in this
case, the values of these probabilities are given solely by the 
corresponding ``beam strengths'' $(K_i + 1 )^2$, with no interference
term.  

In this model, then, we represent particle beam amplitude phases by 
path amplitude phases.  The beam strengths correspond to the number of adjacent paths with the 
same amplitude, and these paths collectively represent the beam.   The model gives a real path ontology according to
which any path with significantly nonzero probability is $d$-close to 
at least one beam.   If the beams themselves are $d$-close, these
probabilities display the familiar quantum interference.   However, if the beams
are widely separated by the $d$ measure, even though they ultimately
recombine at the same point, the probabilities have no interference
term.  The particle follows a path $d$-close to one beam or the
other, with probabilities proportional to the respective beam
strengths. 

\subsubsection{Multiple beam interference}

Consider now a quantum multiple beam interferometry experiment 
in which -- according to the standard intuitive but unrigorous language used
to describe beam interferometry -- a particle leaves a source $A$,
follows one of $n$ linear beam
paths $BP_0 , BP_1 , \ldots BP_{n-1}$ to one of $n$ different slits $S_0
, S_2 , \ldots S_{n-1}$, and then follows a linear path from
the relevant slit to a point $B$ on a detecting screen.   
Suppose that, in a realistic description of the experiment, the beam 
path $BP_i$ has action $S_i = S(BP_i)$, with $\exp ( i S_i ) =
\exp (i \theta_i )$, where the phases $\theta_i$ obey
$0 \leq \theta_i < 2 \pi $, and that the beam along $BP_i$ has strength
$\alpha_i > 0 $, with $ \sum_i \alpha_i^2 = 1$. 

We can extend our two beam model $M2$ to model multiple beams
by introducing multiple regions of 
$K_i$ adjacent paths of constant phase $\exp (i \theta_i )$, where 
$(K_i + 1)$ is proportional to $\alpha_i$.  We obtain the same qualitative
features.   If all beams are $d$-close, our model will reproduce
standard interference; if each pair of beams  is $d$-distant, it
predicts none.  

In particular, to recover the standard quantum
interference predictions, we need to suppose that the beam path separations are microscopic:
$d(BP_i , BP_j ) \approx 0$ for each $i,j$.  
We can achieve this with a generalization of $M2$ (call it $M3$) in which the 
paths $P_1 , \ldots , P_N$ have amplitudes 

\begin{eqnarray}\label{multipath}
A(P_i ) &=& (-1)^{i-1} {\rm~for~} 1 \leq i \leq M_0 -1 \, , \\
A(P_i) &=& \exp ( - i \theta_0 ) {\rm~for~} M_0 \leq i \leq M_0 +K_0 \, , \nonumber \\
A(P_i) &=& (-1)^{i-M_0 -K_0 } {\rm~for~} M_0 +K_0 < i \leq M_1 -1 \, \, , \nonumber \\
A(P_i) &=& \exp ( - i \theta_1 ) {\rm~for~} M_1 \leq i \leq M_1 +K_1 \, , \nonumber \\
A(P_i) &=& (-1)^{i-M_1 -K_1} {\rm~for~} M_1 +K_1 < i \leq M_2 - 1 \, \, ,
\nonumber \\
& \ldots & \nonumber \\
A(P_i) &=& (-1)^{i-M_{n-2} -K_{n-2}} {\rm~for~} M_{n-2} +K_{n-2} < i \leq M_{n-1} - 1 \, \, ,
\nonumber \\
A(P_i) &=& \exp ( - i \theta_{n-1} ) {\rm~for~} M_{n-1} \leq i \leq
M_{n-1} +K_{n-1} \, , \nonumber \\
A(P_i) &=& (-1)^{i-M_{n-1} -K_{n-1}} {\rm~for~} M_{n-1} +K_{n-1} < i \leq N \, \, ,
\nonumber 
\end{eqnarray} 
where we take $M_0$, $M_1 -M_0-K_0$, $M_2 -M_1 - K_1$, 
$\ldots$, $M_{n-1} -M_{n-2} - K_{n-2}$, to be odd 
and $N-M_{n-1} -K_{n-1}$ to be even.
The path amplitudes listed in order from $A_1$ to $A_N$ are
thus now 
\begin{eqnarray}
1, -1 , & \ldots & , 1 , -1 , \exp ( - i \theta_0 )  , \exp ( - i \theta_0 ) , \ldots \exp ( - i \theta_0 ) , 1 , -1 ,
\ldots , \\
& \ldots & , 1 , -1 ,  \exp ( - i \theta_1 )  , \exp ( - i \theta_1 ) ,
\ldots \exp ( - i \theta_1 ) , 1 , -1 ,
\ldots , 1, -1 \,  \nonumber  \\
& \ldots & \nonumber \\ 
& \ldots & , 1 , -1 ,  \exp ( - i \theta_{n-1} )  , \exp ( - i \theta_{n-1} ) ,
\ldots \exp ( - i \theta_{n-1} ) , 1 , -1 ,
\ldots , 1, -1 \,  \, . \nonumber  
\end{eqnarray} 
Here the beam path $BP_j$ is modelled by paths $P_i$ in the range $M_j
\leq i \leq M_j + K_j$.   
We need to take 
$1 < M_0 < M_0+K_0 < M_1 < M_1 + K_1 < \ldots < M_{n-1} +K_{n-1} < N$,
to ensure the beams in our model do not overlap.   
$K_i  \ll N$. 
We also want to take $M_{n-1} + K_{n-1} - M_0 \ll 2D+1$, to ensure the
beam paths are all $d$-close.  
Finally, we take $M_0 \gg 2D+1$ and $N-M_{n-1}-K_{n-1} \gg 2D+1$,
ensuring that there are many more paths ``outside'' the region in 
which the interferometry beams lie than ``inside'' that region. 

Real path quantum theory, via (\ref{pathprob}), then
tells us that the probability distribution of real paths
is dominated by $\approx 2D$ paths around the beam paths, 
each of which has probability proportional to 
$$ (2D )^{-1}  | \sum_i ( K_i + 1 )\exp (- i \theta_i ) |^2   \, .
$$

Note that, if we take this model seriously as a guide to ordinary quantum
interferometry experiments, it suggests the real path in such
experiments is generally very unlikely to be a path lying within any
beam $BP_j$ -- i.e. to be one of the beam paths in our model.
In our model the beam paths are not only all $d$-close to one another,
but considerably closer than this constraint requires.
Our model suggests that a real path can be $d$-close to them while following some exotic non-beam
path through the region in which the beam paths lie, or while going well
outside that region.    

Of course, our toy model relies on many simplifying assumptions.  
A full path integral description
would include infinitely many paths with phases close to $\theta_i$
in the neighbourhood of each $P_i$, and infinitely many more 
exotic paths that are not piecewise linear and have rapidly 
fluctuating phases.   Moreover, even in $1+1$ dimensions, these paths are 
not geometrically related in a way that allows the sort of one
dimensional representation that our toy model and choice of distance function assume.
Nor is it evident which choice of distance function one should make 
for the full set of quantum paths, even for a single particle
travelling between two specified points in space-time.    
These issues clearly ultimately need to be addressed, and we discuss
some of them further below.   
Our present aim, though, is to extract intuitions from and explore the
range of possibilities suggested by discrete toy models.   
We consider next ways of modelling multi-particle 
systems, beginning by considering models of interferometry experiments 
that include the measuring apparatus as well as
the interfering particle.   

\subsection{Real paths in multi-particle configuration space}

We now extend our toy model discussion to consider an interference experiment 
with interfering beams in which the particle is emitted by a source at
$A$ and may arrive at any of various points $B_j$ (for $1 \leq j \leq
l $) on the screen.   

A parsimonious way of doing this is to continue to 
model the experiment by considering only paths of the interfering
particle.   Thus, we could consider sets of paths 
$L_j = \{ P^j_1 , \ldots , P^j_{N_j} \}$ from $A$ to $B_j$,
with corresponding amplitudes of the form (\ref{multipath}), now 
characterized by $j$-dependent parameters $M_k^j , K_k^j , \theta_k^j
, n_j , N_j$ (where $0 \leq k \leq n_j - 1$). 
We can then apply the path probability postulate to the union of these sets
of paths -- i.e. calculate path probabilities conditioned on the
particle arriving at any of the points $B_j$. 
In the regime considered above, with $M^j_{n_j -1} + K^j_{n_j -1 } -
M^j_0 \ll 2 D + 1 , M^j_0 \gg 2D+1 $ and $N_j - M_{n_j - 1 } - K_{n_j
  - 1} \gg 2D+1$ for each $j$, the model reproduces quantum
interference for paths from $A$ to any given $B_j$. 
It gives the 
probability of the particle arriving at $B_j$ as approximately
proportional to $$  | \sum_k ( K_k^j + 1 )\exp (- i \theta_k^j ) |^2
\, , 
$$
and so the quantum predictions for the observed detection ratios 
$\Prob ( A \rightarrow B_j ) / \Prob ( A \rightarrow B_k )$ are recovered, as expected.    

To explore a bit further, we want to extend the model further to
consider paths in the multi-particle configuration space 
of the interfering particle together with constituent particles of the
screen, and continue the model from the time (say $t=0$) at which the particle
is emitted up to some time ($t=T_f$) significantly after the time
($t=T_h$) at which it hits
the screen.   

We take there to be $Z$ constituent particles of the screen.
For the purposes of this model we suppose that, in a quasiclassical
description of the physics, a subset of $Z_j$ of these are appreciably
disturbed by an impact at $B_j$.  We make the further simplifying
assumptions that the possible impact
points $B_j$ are discrete, and the relevant subsets of $Z_j$ disturbed
particles corresponding to distinct $B_j$ are disjoint.   
Rather than considering paths for each individual particle, though,
we first model the screen paths in multi-particle configuration
space by essentially the same toy model considered above.  
Between time $t=0$ and $T_h$, we consider the beam particle and screen
paths separately; after $T_h$ we use a single set of paths to model 
both.  The idea here is that on impact the beam particle becomes part of the screen, 
indistinguishable from the other screen particles: this is not an
essential assumption but simplifies the model.     

We expect screen particles to follow a quasiclassical trajectory at all times,
and we model this by adapting model $M1$.  
Between times $0$ and $T_h$, we take there to be $N'$ possible
configuration space paths in our discrete model of the $Z$-particle 
screen's configuration space, 
listed in order as $P'_1 , P'_2 , \ldots , P'_{N'}$.  
Of these, there are $K'+1$ adjacent
paths with phases $+1$, beginning with $P'_{M'}$, and the rest have alternating phase.  
We suppose there are $N''$ possible configuration space paths
describing the $(Z+1)$ screen particles (now including the absorbed beam
particle) after time $T_h$, listed in order as $P''_1 , P''_2 , \ldots , P''_{N''}$.  
We suppose that the phases of the screen particles' paths after the 
beam particle is absorbed are effectively determined by the absorption point $B_j$
via some interaction Hamiltonian.  Specifically, if we consider paths in which the interferometry particle
arrives at $B_j$, we suppose the $K'' +1$ adjacent screen particle paths 
$P_{M''_j} , P_{M''_j + 1} , \ldots , P_{M''_j + K''}$ to have
phases $+1$, while the other screen particle paths have alternating
phases $\pm 1$.    We assume the quasiclassical
paths corresponding to the distinct absorption points are ordered and
separated by more than $(2D+1)$ from each other and the endpoints, so
that $2D+1 < M''_1 < M''_1 + K'' + 2D +1 < M''_2 < \ldots < M''_{l-1}
+K'' + 2D+1 < M''_l < M''_l + K'' + 2D +1 < N''$. 
We also assume $K'' \ll N''$. 
Note that we take $K''$ to be a constant, 
independent of the impact point $B_j$.  This is to ensure that the probability
of observing an impact at point $B_j$ is not retroactively affected by the 
dynamics of the screen particles after impact.   This is required
so that our model reflects quantum unitarity in the quantum limit 
in which $d(P_i , P_j ) = 0$ for all paths $P_i , P_j$.
Allowing $K''$ to depend on the impact point $B_j$ would mean that 
the post-interaction dynamics of screen particles would 
affect the probabilities of detecting the beam particle at 
given locations on the screen, strongly violating both unitarity and
the no-signalling principle.\footnote{We exclude this possibility
in our models, since we intend the set of allowed paths in our toy
models to reflect physically significant 
features of the quantum path integral as faithfully as possible. 
Our discrete toy models are meant to illustrate plausible
consequences of our real path ontology and path probability postulate
when applied to the full set of paths in realistic models.   Our interest is in
modifying standard quantum theory via the real path ontology and path
probability postulate.   New features that arise solely because of artefacts 
of a particular discrete path integral model are not interesting for
this project.   Although, as we comment later, real path quantum theory
in its present form is not guaranteed to respect the
no-signalling principle, it makes no sense to introduce a further
model-dependent and ad hoc violation.}

The possible complete paths from $t=0$ to $t=T_f$, including an
absorption at 
the screen point $B_j$ at $t=T_h$, 
then take the form 
$$
( \{ {\rm particle~ path}~ P^j_i \} \otimes 
\{ {\rm screen ~path}~ P'_k \} )
\oplus \{ {\rm post-absorption~ screen~ path}~ P''_m \} \} \, ,
$$
where 
$ { 1 \leq i \leq N_j} \,  , {1 \leq k \leq N'} \, ,$ and
$ { 1 \leq m \leq
  N''}$. 

Here $\otimes$ denotes the path in $(Z+1)$-particle configuration
space given by the product of the relevant particle and screen paths,
running from $t=0$ to $t=T_h$, 
and $\oplus$ denotes the composition of a path of this type with
a post-absorption screen path that runs from
$t=T_h$ to $t=T_f$.

We need to define distance functions for products of paths and 
compositions of paths.  
For the moment let us take 
\begin{equation}\label{pathsumproduct}
d ( P \otimes P' , Q \otimes Q' ) = \max ( ( d (P, Q) ,
d(P', Q' ) ) \, , \qquad d (P \oplus Q , P' \oplus Q' ) = \max ( d (P, P' ),
d (Q, Q' )) \, . 
\end{equation}

   This ensures that (as usual neglecting boundary
artefacts
in the model) the non-zero probability paths take the form 
$$
( P_i^j \otimes P_k^{\prime} ) \oplus P_m^{''} \, , 
$$
where 
$P_i^j$ is any nonzero probability particle path in the single-particle
  interferometry 
model with endpoint  $ B_j$, 
$P_k^{\prime}$ is any screen path up to time $T_h$ within $D$  of the
  quasiclassical paths, 
and $P_m^{''}$ is any post-absorption screen path within  $D$  of those
quasiclassical paths that describe the screen after
an absorption at $B_j$. 

It also ensures that we again recover the quantum predictions for 
detection probability ratios 
$$\Prob ( A \rightarrow B_j ) / \Prob ( A \rightarrow B_k )$$ in this
extended model.   

While it is encouraging that we can find a simple model in 
which the one-particle path probability postulate follows from
applying the path probability postulate to a model that 
includes a measuring apparatus or environment (the screen)
as well as the measured particle, the assumptions used in this model
raise many questions.    To list just a few:

Could we obtain similar results
from a more realistic model in which paths for all $(Z+1)$ particles are 
treated on a equal footing?   In which paths live in discretized Galilean or 
Minkowski space-time, with three spatial
dimensions, and the distance function depends (only) on the space-time
geometry?  Or in which we consider the full set of paths in ordinary
continuous Galilean or Minkowski space-time?
Is there a compelling reason to assume that $d ( P \otimes P' , Q \otimes Q' ) = \max ( ( d (P, Q) ,
d(P', Q' )$ and $ d (P \oplus Q , P' \oplus Q' ) = \max ( d (P, P' ),
d (Q, Q' ))$, or are there other interesting options?
What form do we expect the distance function $d$ to take in a realistic
quantum path integral treatment of real systems?   Is it necessarily reasonable
to take the distinct quasiclassical outcomes of measurements to
be necessarily $d$-distant, as we did in our model?   If so, is this
because of the number of particles whose typical 
quasiclassical trajectories are distinct for distinct outcomes, or 
because of the total mass of these particles?  Or is it also relevant
that the quasiclassical trajectories remain distinct indefinitely:
is the length of time (or in the relativistic case, proper path time)
for which paths are distinguishable as (or even
more) relevant as their spatial separation?    
How should particle interactions be handled in real path quantum
theory?   Can we find prescriptions that do not rely on the
approximation in which they are treated by
introducing interaction potentials in the Schrodinger equation 
for each pair of particles, and assuming that the particle number is
fixed?  Could and should the distance function $d(P,Q)$ depend on
interactions -- for instance on the relationships between
the vertices in Feynman diagrams corresponding to paths $P$ and $Q$
-- in a more fundamental field-theoretic treatment?

We believe the first two of these questions can be satisfactorily
(and positively) answered by somewhat more sophisticated toy models.  
As the remaining questions illustrate, though, there is a limit to 
what toy models can persuasively establish.   Fundamental issues
need to be addressed in more realistic settings.  
In what follows we set out some intuitions 
and possibilities to explore, in the hope of both clarifying the vision underlying real path
quantum theory and encouraging wider interest in the research program.  

\section{Possible Path distance functions and their physical implications}

\subsection{Choices of distance function for single particle paths}

What if we apply (\ref{pathprob}) to realistic path integral
models?   A full path integral description of an interferometry
experiment (whether of a microscopic or macroscopic object)
involves uncountably many paths, most of which are not 
close to being the piecewise linear paths that we normally
consider as the interfering beams.  

If real path quantum theory 
does make sense in realistic models, it has one immediately clear empirical
prediction.  Standard quantum interference should be observed 
for interferometry experiments where the beam paths are close, 
as determined by the distance function $d$.  However, quantum
interference should not be observed when the beam paths are 
widely $d$-separated.   For some choices of $d$, these
predictions are very broadly qualitatively similar to those made by dynamical collapse 
models and other intrinsic decoherence models, although the underlying
models are conceptually and ontologically quite different.
For other choices of $d$, the predictions are qualitatively very
different
from those made by any existing intrinsic decoherence model.  

For a microscopic object,
given that the beam paths are microscopically separated from
one another, and that there are infinitely many non-beam paths
between the beam paths, we expect most paths that are microscopically separated
from the beam paths to be non-beam paths.   
{\it If} (\ref{pathprob}) gives 
a well-defined probability measure on the paths,  
we thus expect a very large set $\{ P_{\lambda} \}_{\lambda \in \Lambda}$ of
possible paths from $A$ to $B$, and a probability measure on $\Lambda$
that is roughly uniform on a large subset.  
We also expect that the total probability of finding the 
microscopic object at $B$ is, to very good approximation, 
the quantum probability. 
However, it isn't evident that (\ref{pathprob}) {\it does} give
a well-defined probability measure on paths, for interesting 
choices of the distance $d$.  

At best, then, we get a rather more complicated picture than
toy models suggest, which needs to be worked out carefully,
but which at first sight is not evidently inconsistent 
nor evidently in contradiction with experiment or observation.   
At worst, we have a prescription that as yet makes no rigorous sense.   
However, either way, there are interesting ways to 
alter (\ref{pathprob}) so as to simplify -- and, one might
hope, rigorize -- the picture.   We turn to these next. 

Is there a unique natural choice of $d$?   Even for single
particle paths in Galilean space-time, there seem to be many 
possible candidates.   Consider two paths 
$$ 
P = \{ \, ( x_P (t) , t ) \, : \, 0 \leq t \leq T \, \} \, , 
\qquad 
Q = \{ \, ( x_Q (t) , t ) \, : \, 0 \leq t \leq T \, \} \, , 
$$
for a particle of mass $m$ between points  $A = ( x_P (0 ) , 0 ) = (x_Q (0) , 0 ) $ 
and   $B = ( x_P (T ) , T ) = (x_Q (T) , T ) $.   
Some simple possible definitions of $d$ that capture arguably natural
notions of path separation include
\begin{eqnarray}
d (P , Q ) &=& \max_{ 0 \leq t \leq T} | x_P (t ) - x_Q (t) | \, , \\
d (P , Q ) &=& m \max_{ 0 \leq t \leq T} | x_P (t ) - x_Q (t) | \, \nonumber , \\
d (P , Q ) &=& \int_0^T dt \, | x_P (t ) - x_Q (t) | \, \nonumber , \\
d (P , Q ) &=& T^{-1} \int_0^T dt \, | x_P (t ) - x_Q (t) | \, \nonumber , \\
d (P , Q ) &=& m \int_0^T dt \, | x_P (t ) - x_Q (t) | \, \nonumber , \\
d (P , Q ) &=& ( \int_0^T dt \, | x_P (t ) - x_Q (t) |^2 )^{1/2} \,
\nonumber . 
\end{eqnarray}
One might also explore definitions sensitive to the first and/or higher path
derivatives, for example 
\begin{equation}
d (P , Q ) = \int_{ 0}^{ T} dt \, | x'_P (t ) - x'_Q (t) | \, . 
\end{equation}
Of course, many other options, or combinations of these options, could be explored. 

It seems then, even for thinking about real path quantum theory in the
context of single particle interferometry, that we either need
some new compelling theoretical reason for picking out some particular 
distance function, or empirical guidance.  A theoretical preference 
could perhaps come either from a new theoretical idea purporting 
to explain {\it why} nature might follow the path probability
postulate for some specific choice(s) of distance function, or 
conceivably from establishing that the path probabilities are
rigorously definable only for some specific choice(s).  
But, absent further help from theory, if we take the idea of
real path quantum theory seriously, it seems we need to 
continue exploring empirically whether quantum interference fails, 
and to be open to the possibility that the transition between
interference and effective decoherence could be governed by 
almost any physical parameter or combination of parameters: 
maximum beam separation, average beam separation, separation
time, particle mass, and so on.    
This is no doubt discouraging for those hoping for a 
precise prediction of how and where quantum theory should 
break down.  It may, however, point to the appropriately scientifically
open-minded strategy of 
encouraging experimental tests of quantum
interference in every possibly interesting new physical parameter range.  
If real path quantum theory is relatively theoretically underconstrained,
by the same token it points to previously unconsidered theoretical possibilities.
It suggests our theoretical understanding of the range of plausible empirical implications 
of unified models of quasiclassical and quantum physics has been too
limited and should be broadened.  

\subsection{Many particles and composition rules}

These comments apply even more strongly when considering 
multi-particle systems.   
We have already noted interesting choices of distance function
that violate the path sum rule 
$$
 d (P \oplus Q , P' \oplus Q' ) = \max ( d (P, P' ),
d (Q, Q' )) \, . 
$$
proposed in (\ref{pathsumproduct}). 
The path product rule
$$
d ( P \otimes P' , Q \otimes Q' ) = \max ( ( d (P, Q) ,
d(P', Q' ) \, , \qquad
$$
is equally open to question.   
Other seemingly mathematically natural possibilities for $n$
distinguishable
particles include
\begin{eqnarray}
d ( P_1 \otimes P_2 \otimes \ldots \otimes P_n  , Q_1 \otimes Q_2
\otimes \ldots \otimes Q_n ) &=&  \sum_i d (P_i , Q_i ) 
\, , \qquad \\
d ( P_1 \otimes P_2 \otimes \ldots \otimes P_n  , Q_1 \otimes Q_2
\, \otimes \ldots \otimes Q_n ) &=&  \frac{1}{n} \sum_i d (P_i , Q_i ) 
, \qquad \nonumber \\
d ( P_1 \otimes P_2 \otimes \ldots \otimes P_n  , Q_1 \otimes Q_2
\otimes \ldots \otimes Q_n ) &=&   ( \prod_i d (P_i , Q_i ) )^{1/n}
\, , \nonumber 
\end{eqnarray}
and of course many others could be considered.  

Even more generally, even for distinguishable particles,
in principle the distance function for a product of 
paths need not be expressible as a function of individual
path distance functions at all. 

For indistinguishable particles, clearly, the distance function should
respect the permutation symmetry.   One possibility would be 
to frame a definition of a symmetric distance function $d_{\it symm}$
in terms of one of the distance functions above:
$$
d_{\rm symm} ( P_1 \otimes P_2 \otimes \ldots \otimes P_n  , Q_1 \otimes Q_2
\otimes \ldots \otimes Q_n ) =
\min_{\rho} 
d ( P_1 \otimes P_2 \otimes \ldots \otimes P_n  , Q_{\rho(1)} \otimes Q_{\rho(2)}
\otimes \ldots \otimes Q_{\rho(n)} ) \, , 
$$
where the minimum is taken over all permutations $\rho$.

One further significant theoretical constraint here arises from the fact
that, if real path quantum theory is fundamentally correct, it 
should in principle be applied to the entire universe, while
if it is to be of any empirical use, it must be applicable to 
small subsystems.  An effective $d$ for single or few particle
subsystems must be derivable from the definition of $d$ 
for a many-particle system, even if the latter is not directly
defined in terms of the former.  But this still leaves many
possibilities.  

\subsection{Lorentz covariant rules for paths in Minkowski space}

A major reason for optimism that the quantum path integral may be 
fundamental is that -- formally at least --
it can naturally incorporate Lorentz and other
symmetries.   Encouragingly for real path quantum theory,  
we can also find relatively simple and seemingly 
natural Lorentz invariant measures of distance
defined on reasonably general classes of paths.  

Consider points $A$ and $B$ in Minkowski space, where $B$ is in the
causal future of $A$.  Let $S_A$ and $S_B$ be spacelike hyperplanes
through $A$ and $B$ respectively.   

Let $P = ( X_P ( \lambda ) : 0 \leq \lambda \leq 1 )$ 
be a  parametrised path in Minkowski space between points $A$ and 
$B$, where $X_P ( 0 ) = A$, $X_P (1) = B$, and the four-vector $X_P$ is a continuous
function of $\lambda$.  
Take the Minkowski metric with the convention that spacelike
vectors have positive length and $c=1$, i.e.
$\Delta ( x, y, z, t ) = x^2 + y^2 + z^2 - t^2$. 
We say $P$ is a {\it causal} path if $X_P (\lambda ' ) $ is in the 
causal future of $X_P ( \lambda )$ whenever $\lambda' > \lambda$. 
Clearly, if $P$ is causal, it lies between $S_A$ and $S_B$. 
We say $P$ is {\it non-causal} otherwise.   
We say $P$ is {\it anti-causal} if there exists $ \lambda' > \lambda$
such that $X_P ( \lambda' )$ is in the causal past of $X_P (\lambda
)$. 

Let $Q$ be a path that is not anti-causal (but 
not necessarily causal) path also lying between
$S_A$ and $S_B$. 

One simple candidate measure of distance for two such paths $P,Q$ between
points $A$ and $B$ in Minkowski space-time is 
$$
d_1 ( P , Q ) = \max_{\lambda , \lambda' } ( \Delta ( x_P (\lambda ) -
x_Q ( \lambda' ) )) \, ,
$$
the maximum spacelike separation between any pair of points on the
two paths.   

Note that if $Q$ is not causal, it includes two spacelike
separated points, and so if we extended this definition to
pairs of non-causal paths we would have $d_1(Q,Q) >
0$, i.e. $d_1$ would violate the identity of indiscernibles.   
The distance function thus distinguishes causal paths $P$,
for which $d_1 (P,P) = 0$, from non-causal paths $Q$, for which
$d_1 (Q,Q)>0$.  
Given this distinction, one arguably natural prescription for real path quantum theory
in Minkowski space is then to allow only causal paths to be realised, while 
allowing amplitudes from non-causal but not anticausal paths to 
contribute to their probabilities via 
the path probability postulate (\ref{pathprob}).  
(Anticausal paths are ignored altogether in this prescription.) 
This would have the intuitively satisfactory consequence that only
causal paths can be physically realised.   
It is also interesting to explore whether allowing
all paths to be realised has physically sensible consequences,
i.e. whether the fact that $d_1 (Q,Q)>0$ 
for non-causal paths $Q$, and the highly oscillatory variation of
the phase in the neighbourhood of such paths, in any case suppresses the probability of highly
non-causal paths being realised.  

Another interesting Lorentz invariant distance function for pairs
of paths $P,Q$ with $P$ causal and $Q$ not anti-causal, 
is 
\begin{equation}
d_2 (P, Q) = \int_P \, d \tau (\lambda) \max_{\lambda'} \Delta ( x_P (\lambda )
- x_Q ( \lambda ' ) ) \, , 
\end{equation}
where $\tau ( \lambda )$ is the proper time along $P$ from $A$ to 
$ x_P ( \lambda )$.   This measure is sensitive not only to the 
space-like separations between points on the paths, but to the 
the proper time interval along $P$ for there is any given space-like 
separation between $P$ and $Q$.  Note that in this form this measure
is not defined if $P$ is not causal, and is not symmetric for causal
$P$ and $Q$.   It also does not satisfy the identity of
indiscernibles: if $P$ is an everywhere null causal path, then we have 
$d_2 (P,Q) = 0$ for all paths $Q$.   

(The definition could, of course,
be extended in various ways.  For example, for non-causal but not anti-causal $P,Q$, we could define 
\begin{equation}
d_2^{\prime} (P, Q) = \int_P^{\prime} \, d \tau (\lambda) \max_{\lambda'} \Delta ( x_P^{\prime} (\lambda )
- x_Q ( \lambda ' ) ) \, , 
\end{equation}
where $P^{\prime}$ is the not necessarily connected sub-path of $P$ that is
the maximal sub-path comprising causal segments, and $\tau$ is now
the proper time along each such segment.  
We could also symmetrise the definition by hand: 
\begin{equation}
d_2^{''} (P, Q) =  \frac{1}{2} ( d' (P, Q ) + d' (Q, P ) ) \, .  
\end{equation}
As in the Galilean case, we could generalise in other ways too,
for example by using monotonically increasing functions of 
$\Delta ( x_P (\lambda )
- x_Q ( \lambda ' ) )$ in 
$d_1$ or $d_2$, by defining a proper time average version of $d_2$ for
non-null causal paths $P$ by taking $\frac{1}{\tau_P} d_2$ (where
$\tau_P$ is the total proper time along $P$) and so forth.

We stress that, as in the Galilean case, the suggestion 
here is not that any one of these distance functions (or some combination), or the path probability
postulate (\ref{pathprob}), or the prescription that only causal
paths can be realised, must necessarily be right.  
There are many other possibilities.  
What we find encouraging is 
that the existence of Lorentz invariant
distance functions transforms a conceptual problem (is there
any conceivable realist Lorentz invariant 
generalisation of quantum theory?) into a technical problem
(can we find a well-defined version of real path quantum theory
for some choice(s) of Lorentz invariant distance function and path
probability postulate?).   
Although the technical problem is formidable and we are far from
understanding whether it is solvable, the existence of Lorentz invariant path distance 
functions suggests that there is no {\it purely 
conceptual} no-go result preventing the possibility 
of Lorentz invariant path-based solutions to the quantum reality problem.  

\subsubsection{Generally covariant path distance functions}

We can also find relatively simple generally covariant definitions
of the distance for a reasonably general class of paths in fixed Lorentzian background space-times
other than Minkowski space.    
For simplicity we focus here on space-times with no closed 
time-like curves and trivial spatial topology. 

For two non-causally separated points $X, Y$ in
such a space-time we define a distance function, $\tilde{\Delta} (X, Y)$ to be the minimum
spacelike distance along a space-like geodesic from
$X$ to $Y$. 

Now consider points $A$ and $B$ in our spacetime, where $B$ is in the
causal future of $A$.  Let $S_A$ and $S_B$ be spacelike hypersurfaces
through $A$ and $B$ respectively, with the property that 
their proper time separation is bounded. 
That is, if we define $\tau(X,Y)$ to be the maximum proper time 
along any causal path from a point $X$ to some point $Y$ in its 
causal future, we have $\tau(X,Y) \leq \tau_0 $ for all $ X \in S$ and
$Y \in S'$. 

Now, defining causal and anti-causal paths as before 
let $P$ be a causal path between $A$ and $B$, which (necessarily)
lies between $S_A$ and $S_B$, and let $Q$ be a path that is not
anti-causal (but not necessarily causal) between $A$ and $B$ that
also lies between $S_A$ and $S_B$.  

We can define 
$$
d( P , Q ) = \max_{\lambda , \lambda' } ( \tilde{\Delta} ( x_P (\lambda ) , 
x_Q ( \lambda' ) )) \, ,
$$
where as before $\lambda$, $\lambda'$ define parametrisations of
$P,Q$ respectively and we use the definition of  
$ \tilde{\Delta} ( X, Y )$ just given. 

As in the Minkowski case, extending this definition to pairs of
non-causal
paths would imply $d(Q,Q)>0$ for  
non-causal $Q$.  Again, one
arguably natural option for real path quantum theory
is to postulate that only causal paths can be realised, while 
allowing amplitudes from non-causal but not anticausal paths to 
contribute to their probabilities via 
the path probability postulate (\ref{pathprob}).  

The other definitions considered in the Minkowski case can similarly
be extended. 
Once again we stress that, as for the Galilean and Minkowski space cases,
we presently see no compelling reason for singling out any of these particular definitions
of covariant path distance or real path probability prescriptions: the encouraging
point is that covariant definitions exist.   

\subsection{Suppressing unphysical paths}

``Pathological'' paths -- which may traverse many regions of space-time
very far from the stationary path and from each other, and may be
very rapidly varying or even undifferentiable -- are problematic in standard path integral
quantum theory.  Technically, they make it hard to define a computable
path integral in realistic models.   Physically, they seem to make it
hard to assign a sensible meaning to the path integral, even at the
level of intuition: does one really want to say that the real behaviour 
of a system is, in some sense, dominated by pathological paths?  

The path probability postulate (\ref{pathprob}), together with 
the scope for choices of the distance function, offer some hope
of ameliorating or even eliminating these problems.   
A distance function sensitive to spatial separation would mean that
a pathological path that travels far from the stationary paths 
makes little contribution to the probability of the latter being
realised.
Distance functions sensitive to first or higher derivatives can
also suppress the contributions of rapidly varying or undifferentiable
paths. 
If the intuition that the Feynman-Hibbs argument can be rigorized
in real path quantum theory is justified, then 
the rapid oscillations of phase of paths in the neighbourhood
of pathological paths tend to suppress the probability of 
these paths being realised themselves, since their probabilities
will be close to zero.   

At the risk of multiplying hypotheses, it is worth mentioning 
that another strategy for suppressing the probabilities and
amplitudes of undesirably pathological paths could also be 
used.   Real path quantum theory is an example of -- in Bell's
terminology -- a beable theory, that is, a theory with a 
sample space of possible realised ontologies (in this case,
possible paths in the appropriate configuration space) and a probability
distribution on that sample space.  
A quite general way \cite{kent2013beable} of producing potentially interesting
generalisations of a beable ontology is to allow the probability of any given 
configuration of ``beables'' being realised to depend, via 
simple rules, directly on the properties of that configuration,
as well as on the Hamiltonian and boundary conditions of the
underlying quantum theory.    

In the case of real path quantum theory, one could further modify
the path probability postulate (\ref{pathprob}), for example by
adding a prefactor in the form of a non-negative weight function $w(P)$ 
that suppresses the probability of pathological paths $P$ being
realised.   Thus, we could take
\begin{equation}\label{pathprobweight}
\Prob (P) = C' w(P) \, | \int dQ \exp (- i S(Q) )  \exp ( - d (P, Q))
|^2  \, ( \int dQ \exp ( - d (P, Q) ))^{-1}  \, , 
\end{equation}
where $w(P) \geq 0$ is significantly nonzero for physically
reasonable paths and small for pathological paths, 
and the constant $C'$ normalizes this new probability distribution. 
(We have already considered a version of this in suppressing
non-causal paths in Lorentzian space-times.) 
In principle, one can consider any choice of $w(P)$: for example,
it could depend on the maximum or typical curvature of the path $P$,
or on some measure(s) of its derivative(s).   In Galilean or 
Minkowski space-time, one could even ensure that any realised $P$ must
be piecewise linear with a characteristic linear scale, by setting $w(P) = 0$
unless $P$ comprises linear segments of given length $\delta$ 
or given proper time $\delta \tau$, if one wished to introduce 
new fundamental scales into the ontology.  

\subsection{Comments on double ontologies and the problem of tails}

Superficially, real path quantum theory (in models where it is 
rigorously defined) has some resemblance to 
de Broglie-Bohm theory.   In the simplest scenario, that of models of $N$ distinguishable
particles, both approaches produce a probability distribution on
trajectories in $N$ particle configuration space.    

In fact, there are major differences.  The probability distributions
are different, for a given initial quantum state and Hamiltonian.
Real path quantum theory generally predicts different physical outcomes from 
those of standard quantum theory, whereas standard de Broglie-Bohm
theory predicts the same outcomes.  The basic postulates of real path
quantum theory extend naturally to relativistic settings, whereas
de Broglie-Bohm theory is hard to relativise.    

There is also a significant difference in the ontologies.  
Standard de Broglie-Bohm theory offers the possibility of 
two separate although related ontologies, defined by taking
the beables to be respectively the quantum wave function (or some derivative
thereof, such as the mass density) and the de Broglie-Bohm   
trajectories.  That the first of these choices is identical to
the one proposed (albeit in many different versions with many different
interpretative strategies) by many Everettians lead
to Deutsch's well-known charge\cite{deutsch1996comment} that
``pilot-wave theories are parallel-universe theories in a state of
chronic denial''.   Of course, one logically consistent defence 
for de Broglie-Bohm theorists is observe that in any physical theory
one has to make choices about which parts of the mathematical
formalism are beables -- i.e. define the ontology -- and which
parts are auxiliary, and to decide to declare by fiat that the
particle trajectories are beables while the wave function is not. 

Whether one is entirely happy with this defence is ultimately
a matter of metaphysical taste.      
Since in practice many are not, it seems worth noting that real path quantum
theory does not seem vulnerable to the same charge of a potential
or actual double ontology.   Paths are fundamental in the formulation 
of this approach to quantum theory.  It produces a probability
distribution on paths.  The ontology is as expected from any
straightforwardly probabilistic physical theory: one element of
the sample space (in this case one path) is randomly chosen from
the probability distribution, and it is (only) this element that 
is physically realised.     The wave function is not an ontological
competitor.  It plays no fundamental
role, emerging only -- and only in some 
situations -- as an approximately defined quantity that gives a convenient
alternative approximate mathematical description of expected
experimental outcomes.   
  
Real path quantum theory also has some resemblance to GRWP and other dynamical
collapse models \cite{ghirardi1986unified,GPR,diosi}.   Like these models, it can preserve
quantum interference for suitably microscopic interference
experiments, while predicting an intrinsic decoherence that suppresses
interference for suitably macroscopic interference experiments.  
Collapse models, like de Broglie-Bohm models, offer the possibility
of at least two distinct types of ontology.  One is the ``flash ontology'' originally
defined by Bell for the discrete GRW model
\cite{bell2001there,kent1989quantum,tumulka2006relativistic},
or its analogue for continuous dynamical collapse models. 
The other is some ontology defined by or derived
from the collapse model wave function (again, for example, via the
mass density).   

Worries about a double ontology perhaps carry less 
force for GRW models than for de Broglie-Bohm theory, for various
reasons.  First, some might perhaps argue that the ``flash ontology''
is less compellingly natural than the trajectory ontology. 
Second, the two ontologies are in any case roughly aligned --
in the sense that, at least on a (perhaps overly) superficial reading
they tell similar stories.   In a standard quantum measurement
setting, the collapse model wave function tends towards a 
description of system and apparatus corresponding to one 
measurement outcome, while the component corresponding to
the other outcome is swiftly and exponentially suppressed;
similarly, the flashes congregate around apparatus particle
position locations corresponding to one outcome, and 
swiftly and with increasingly high probability tend to
avoid the other.  There is no straightforwardly Everettian
competitor ontology here.    

Nonetheless, these double ontology worries are not entirely eliminated.   
In particular, a feature that many have found problematic 
is that, although the components of the wave function 
corresponding to ``unselected'' outcomes decay rapidly
after measurement, they never entirely disappear, and the 
physical description of the alternative outcome remains
encoded in these small but nonvanishing components.
This ``problem of tails'' opens up the possibility of 
a subtle but persistent residual Everettian ontology.
Whether it is a potential concern depends on whether one is 
justified in neglecting small amplitude components of
the wave function as essentially irrelevant.  However,
this very issue lies at the heart of the problem of probability
in many-worlds quantum theory \cite{geroch1984everett,kentoneworld,albert}.  Those (many) who believe
the problem of probability has no satisfactory solution
thus find it hard to dismiss the problem of tails completely.    
On the other hand, those (also many) who believe the problem of probability
in many-worlds quantum theory does have a satisfactory
solution tend also to take many-worlds quantum theory as a 
satisfactory answer to the quantum reality problem, and
are correspondingly less motivated to take dynamical
collapse models seriously in the first place.    

Again, then, it seems worth noting that real path quantum 
theory does not suffer from a problem of tails.   The 
wave function plays no fundamental role.   One path is
randomly chosen to be physically realised.   The other
paths are not chosen, and so do not form part of the
ontology.   

\section{Discussion} 

All approaches to quantum theory and its generalisations -- whether or not they are motivated
by the quantum reality problem -- currently have deep problems,
if not in the eyes of their proponents, then certainly in the view
of most dispassionate outsiders.   This makes the task of the theorist
interested in these fundamental questions very challenging.  We have to strive for conceptual
clarity and mathematical rigour. 
We have to try to understand 
which approaches might at least potentially ultimately produce 
a mathematically and conceptually complete 
description of nature, and which are likely dead ends that are
bound to fall short of this goal.   
At the same time, so long as no approach offers a fully satisfactory
solution, we should
not neglect potentially useful insights that incompletely developed
models can provide.   To list some examples: 

Could it be literally true that unitary quantum 
theory is fundamental and the wave function represents reality, as
Everett first suggested?   Can we even make conceptual sense of
the idea and recover ordinary science?   Many physicists think not, but few
would deny that the idea has been a theoretically and practically
useful way of thinking about quantum theory and quantum
information theory, even if it is best thought of as an unrealistic
limiting case.  In particular, the idea makes a clear prediction
that could not so unambiguously be extracted from pre-Everettian
quantum theory. Namely that, however large and complex a physical
system may be, in principle, provided it can be suitably controlled
while effectively isolated from the environment, it will display
quantum interference.   

Is wave function collapse a real and fundamental physical phenomenon?   
Perhaps, perhaps not.   But the idea has motivated the theoretically
and scientifically fertile dynamical collapse model program, including
speculative but intriguing ideas relating collapse to gravity \cite{diosi,penrose}. 
Again, these ideas suggest clear predictions: that quantum 
interference {\it will} fail in a regime where a well-defined
collapse model suggests that the superposed states collapse 
to a component.  

Is Bell's notion of beables \cite{bell1976theory, bell1987beables} a good way of thinking about fundamental
physics, and the right language in which to address the quantum
reality problem?   Again, many, perhaps most, physicists presently think not.
But nonetheless the idea motivates new generalisations of quantum
theory \cite{valentini1991signal,kent1997beyond,kent2013beable}, and new ways of thinking about the relationship of
quantum theory and gravity, that can be scientifically useful 
and suggest new experimental tests -- spin-offs which might
lead to new physics, even if that physics turns out not to 
be written in the language of beables. 

We would thus tentatively suggest that, while further development is of course 
urgently needed, real path quantum theory has some scientific yield,
even in its presently undeveloped
state.   
This is the motivation to explore empirically whether quantum interference fails, 
with an open mind about what governs the transition between
interference and effective decoherence.   If we take the idea
of real path quantum theory seriously, then until further theoretical
and/or mathematical constraints are uncovered, this transition could be governed by 
almost any physically interesting parameter or combination of parameters: 
maximum beam separation, average beam separation, separation
time, particle mass, and so on.    

As we remarked earlier, in one important sense this is scientifically
quite discouraging.  
In an ideal universe, we might prefer any idea for a generalisation of quantum theory to be strongly
constrained (or, even better, to offer a unique alternative) 
and to make precise predictions of how and where quantum theory should 
break down.  It would be pleasing to identify some critical, and
preferably soon feasible, experiment that distinguishes between
quantum theory and the generalisation, so that we can refute or
confirm the idea once and for all.  

However, we do not get to choose the universe we
live in, nor the size or structure of the class of mathematically consistent
theories that generalise our best current theories of that universe. 
Our theoretical understanding of the class of unified models of
quasiclassical and quantum physics 
and the range of plausible empirical implications of these models may
indeed -- as the idea of real path quantum theory presently suggests -- have been too
limited.  It appears there may be a very large class of consistent
generalisations of quantum theory (as Refs. \cite{kent1997beyond,kent2013beable} also suggest,
using different ideas and different classes of generalisations from
those considered here).    
Unless and until our theoretical understanding advances further, it may be -- 
as the preliminary version of real path quantum theory presented here suggests -- that the appropriately scientifically
open-minded and curious strategy is to test quantum 
interference experimentally in every possibly interesting new physical
parameter range, rather than -- for example -- focussing only on those
picked out by simple dynamical collapse models, well-motivated though
tests of dynamical collapse models certainly are.   

One significant theoretical constraint whose implications for real
path quantum theory need to be properly understood is the
no-signalling principle.  Practically speaking, for models
in which a physically sensible multi-particle or field configuration distance
function allows real path quantum theory for small subsystems
of the universe to be approximately derived from a fundamental
real path quantum theory for the universe, we do not expect 
agents' choices about actions on one small subsystem 
to have a significant effect on the real path probability 
distribution of another distant small subsystem.  
So, for such models, we expect no-signalling to
be an excellent approximation by this operational
definition.   However, the possibility of enforcing no-signalling 
precisely and in all circumstances, and the implications
of such a constraint, needs further investigation.
A failure of no-signalling need not be a reason for rejecting
any form of real path quantum theory, particularly if the
observational consequences are negligible.  There
are consistent Lorentz invariant generalisations of quantum 
theory that in principle allow superluminal signalling but
evade causal paradoxes \cite{akcausality}.  However, the
no-signalling principle remains, at the very least, a
very interesting theoretical
constraint on generalizations of quantum theory \cite{gisin1990weinberg,polchinski1991weinberg}. 

The ultimate vision of those who take path integral quantum theory as
fundamental to all of physics is a path integral formulation of
quantum gravity and quantum cosmology.   It hardly needs saying that it would be very
interesting to explore the possibilities opened up by real path
quantum theory in this context, the range of potentially natural
distance functions between cosmological paths, and the possibility 
of using the path probability postulate and the properties of distance
functions to allow rigorous definitions and calculations.        

Finally, we wish to emphasize that the appeal and generality of the path integral
formalism and its consequent possible value in generalising quantum theory
and/or addressing the quantum reality problem has of course also been
recognised by others.  In particular -- to list just those of
which we are aware -- Sorkin
\cite{sorkin1995quantum,sorkin2007quantum}, 
Dowker-Johnston-Sorkin \cite{dowker2010hilbert}, Gell-Mann and Hartle
\cite{hartle2008quantum,gell2012decoherent}, Stamp
\cite{stamp2012environmental}, and Spekkens \cite{spekkens} have ongoing 
research programs exploring these questions from various
perspectives.   The independent work described here starts
from different premises, and as far as we presently understand
things it seems to suggest a different
research agenda and qualitatively (as well as quantitatively) different ontological and experimental
conclusions.   It would, however, certainly be interesting to explore
possible connections between, or combinations of ideas from, these
programs.    

\section{Acknowledgements}
This work was partially supported by a Leverhulme Research Fellowship, a grant
from the John Templeton Foundation, and by Perimeter Institute for Theoretical
Physics. Research at Perimeter Institute is supported by the Government of Canada through Industry Canada and
by the Province of Ontario through the Ministry of Research and
Innovation.
I thank Bei Lok Hu for a helpful conversation.  
\section*{References}

\bibliographystyle{plain}
\bibliography{pathreality}{}

\end{document}